\documentstyle[aps,prl,epsfig,twocolumn]{revtex}
\begin{document}

{\bf Reply to the Comment 
``Intercluster correlation in Seismicity" by Helmstetter and Sornette}

\bigskip
%

The comment by Helmstetter and Sornette \cite{com} addresses an 
important point of our paper \cite{ours} that needs to be clarified.
Any scaling measure, including also the diffusion entropy (DE) method, 
when applied to the earthquake time series, yields anomalous scaling 
for two reasons, long-range time correlation  (intermittency) and
the Pareto's statistics. 
In fact, the DE random walker position is determined by the 
number of events (earthquakes) in a given time interval, 
which, in turns, is related to 
both  the number of clusters and  the cluster size. 
The distribution of the time distances between clusters, 
$\psi(\tau^{[m]})$ in ref.\cite{ours}, 
determines the number of  clusters,  
while the Pareto's law 
\begin{equation}
p(n)\propto n^{-\gamma}
\end{equation}
gives  the probability to get  a cluster of size $n$.
The resulting asymptotic scaling detected by the 
DE is the scaling of the most 
anomalous of the two processes, i.e. the
process with the lowest power law exponent.   

The authors of the comment \cite{com} discuss 
the Generalized Poisson (GP) model  with an exponent $\gamma=2.25$ \cite{note},
which generates $\delta=0.8$. In fact, according to the Continous
Time Random Walk theory \cite{montroll},
the scaling is $\delta=1/(\gamma-1)$
if $2<\gamma<3$, $\delta=0.5$ for $\gamma>3$
and $\delta=1$ for $\gamma<2$.

%
\begin{figure}[htbp]
	\begin{center}
		\includegraphics[width=2.8in]{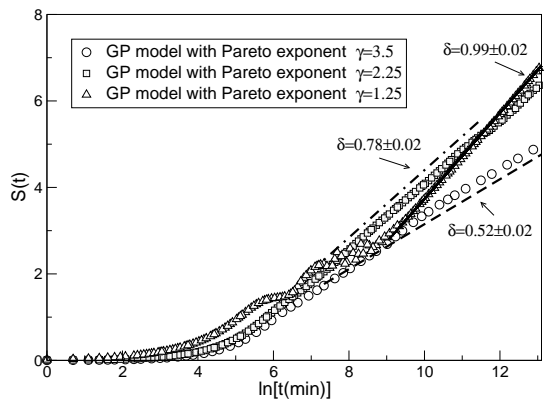}
\vskip 0.2 cm
		\caption{ 
We plot the DE analysis for different synthetic series generated 
with the GP model and three different exponents of the Pareto law, i.e. 
$\gamma=3.5,2.25,1.25$. The values of $\delta$ obtained by fitting the 
asymtotic behavior are also reported with the corresponding error. These values increase
as  $\gamma$ diminishes  passing  respectively from 
$0.52 \pm 0.02$ (for  $\gamma=3.5$)
to $0.78 \pm 0.02$ (for  $\gamma=2.25$) 
and finally to  $0.99 \pm 0.02$ (for  $\gamma=1.25$). 
}
	\end{center}
	\label{fig:figura1}
\end{figure}

We have repeated the calculations with $\gamma=2.25$ and 
the results, shown  in fig.1, confirm the theoretical prediction. 
In the figure we plot also two other cases in order to illustrate how  
$\delta$ changes as a function of  $\gamma$. 
In Ref.\cite{ours} we explored the range $\gamma=2.5 - 3$.
 
The experimental data yield a value $\delta=0.94$ 
confirming that $\psi(\tau^{[m]})$ has a power law 
tail with an  exponent $\mu = 2.06$.
Moreover, the fact that the scaling observed 
in fig. 2 of our previous paper \cite{ours}
is not  influenced by the threshold of magnitude considered, is an additional
evidence that what we are actually observing is the real scaling 
determined by the  clusters distances.

The LR model \cite{ours} is a
minimal generalization of the GP model 
(only  $\psi(\tau^{[m]})$ changes).
Therefore it is very interesting that a more sophisticated
model, like the ETAS, which  reproduces also other
properties of seismicity, yields the same $\delta$
observed in the LR and in the experimental data. 
However we would like  to point out  that we define
 $\tau^{[m]}$ as the distance between two consecutive clusters,
while in \cite{com} this symbol refers to distance between 
two consecutive earthquakes 
with magnitude greater than $M=5$. So there is no contraddiction
between the correlations found in \cite{com} and our results.

\bigskip

{Mirko S. Mega$^{1}$, Paolo Allegrini$^2$,
Paolo Grigolini$^{1,3,4}$, Vito Latora$^{5}$,
Luigi Palatella$^{1}$, Andrea Rapisarda$^{5}$ and Sergio
Vinciguerra$^{5,6}$}
\vskip 1 cm

{$^{1}$Dipartimento di Fisica
dell'Universit\`a di Pisa and INFM, via Buonarroti 2, 56127 Pisa, Italy}

{$^2$Istituto di Linguistica Computazionale del CNR,
Area della Ricerca di Pisa,
Via G. Moruzzi 1, 56124, Pisa, Italy}

{$^{3}$Center for Nonlinear Science, University of North
Texas,   P.O. Box 311427, Denton, Texas 76203-1427 }

{$^4$ Istituto di Biofisica del CNR, Area della Ricerca di
Pisa, Via Alfieri 1, San Cataldo,
56010, Ghezzano-Pisa, Italy}

{$^{5}$ Dipartimento di Fisica e Astronomia,
Universit\`a di Catania, and INFN sezione di Catania,
Via S. Sofia 64, 95123 Catania, Italy}

$^{6}$ Osservatorio Vesuviano - INGV, Via Diocleziano 328,
80124 Napoli, Italy

\bigskip
\noindent



\begin{thebibliography}{10}
\bibitem{note} In the literature Pareto's law is 
referred as the cumulative distribution. Our exponent
$\gamma$ is related to the exponent reported in ref.\cite{com} 
by the expression $\gamma=1+b/\alpha$.
\end{thebibliography}
\end{document}